# Highly Efficient Acoustooptic Diffraction in $Sn_2P_2S_6$ Crystals

Martynyuk-Lototska I.Yu.[1], Mys O.G.[1], Grabar A.A.[2], Stoika I.M.,[2] Vysochanskii Yu.M.[2] and Vlokh R.O.[1]

[1] Institute of Physical Optics, 23 Dragomanov St., 79005 Lviv, Ukraine, e-mail: vlokh@ifo.lviv.ua
[2] Institute for Solid State Physics and Chemistry of Uzhgorod National University, 54 Voloshyn St., 88000 Uzhgorod, Ukraine, e-mail: agrabar@univ.uzhgorod.ua

**Abstract**

We have studied the acoustooptic (AO) diffraction in $Sn_2P_2S_6$ crystals and found that they manifest high values of AO figure of merit. The above crystals may therefore be used as highly efficient materials in different AO applications.

## Introduction

Tin thiohypodiphosphate crystals ($Sn_2P_2S_6$) are wide-bandgap semiconductor crystals with a proper second-order paraelectric-to-ferroelectric phase transition $2/m \leftrightarrow m$ at $T_c = 337\,K$ [1]. These crystals are transparent in a wide spectral range (from $\lambda = 0.53\,\mu m$ to $\lambda = 8.0\,\mu m$ [2]), have high enough values of electrooptic coefficients ($r_{11} = 1.74 \times 10^{-10}$ m/V at the room temperature and the light wavelength $\lambda = 633\,nm$ [2,3]) and manifest photorefractive properties [4,5]. This is why $Sn_2P_2S_6$ crystals can be used in optoelectronics, in particular in the devices for operating optical radiation. In our recent papers [6,7] we have estimated acoustooptic figure of merit (AOFM) $M_2$ of $Sn_2P_2S_6$ crystal, basing on the measured ultrasonic velocities ($V = 1560\,m/s$), piezooptic coefficients ($\pi \approx 7\,pm^2/N$) and the other available data. A very high value ($M_2 = 2 \times 10^{-12}\,s^3/kg^2$) has been so obtained. This would insure a great potential of this material for acoustooptic (AO) devices. Since the experimental data on AO diffraction in $Sn_2P_2S_6$ are still absent in the literature, the present paper is devoted to direct measurements of the corresponding parameters.

## Experimental

$Sn_2P_2S_6$ single crystals were grown using a conventional vapour-transport method [8]. Single crystalline samples were oriented with X-ray diffraction method and prepared in the shape of parallelepipeds. Since $Sn_2P_2S_6$ crystals belong to monoclinic point group *m*, orientation of their optical indicatrix in the plane of symmetry (010) depends on the light wavelength, temperature, etc. According to adopted conventions, the Y axis of the coordinate system for $Sn_2P_2S_6$ crystals is orthogonal to the symmetry plane and the X axis is directed almost along the spontaneous polarization vector [3]. Hence, only Y axis of the crystallographic coordinate system coincides with the corresponding principal axis of the optical indicatrix, while X and Z axes at the room temperature and the light wavelength of 632.8 nm are rotated by $\simeq 45°$ in the XZ plane with respect to the crystallographic directions.

All the measurements were performed with single-domain samples. To reach a uniformly poled state, the samples were heated above the phase transition point ($\sim 370$ K), annealed at ~350 K about 1 h and then slowly cooled down under the external electric field ($\sim 600$ V/cm) applied along the X axis.

The experimental studies of AO diffraction in $Sn_2P_2S_6$ crystals were carried out under the conditions of propagation of transverse acoustic wave along the direction [010], its polarization along [101] (the acoustic wave frequency was 26.5 MHz). Incident circularly polarized optical radiation of He-Ne laser propagated parallel to the direction [101]. The ultrasonic wave was excited using a piezoelectric $LiNbO_3$ transducer. The intensity of the incident light $I_i$ and the intensity of the zero-order diffraction maximum $I_0$ were measured with semiconductor photodetector. The diffraction efficiency $\eta$ was calculated according to the relation

$$\eta = \frac{I_i - I_0}{I_i}, \qquad (1)$$

where $I_i$ and $I_0$ values were measured as functions of the drive electric power applied to the piezoelectric transducer.

**Results and Discussion**

As the result of acoustooptic interaction the 0 and ±1 diffraction maximums appeared. The results of experimental measurements of the AO diffraction efficiency for the case of interaction of the optical wave with the transverse acoustic wave in $Sn_2P_2S_6$ crystals are presented in Figure 1. The AO diffraction efficiency reaches the value $\eta = 16\%$ if the power of the applied electric signal is equal to $P_{el} = 1.56$ W. Notice that this value does not represent a limit of some kind and it is possible to expect further increase in the diffraction efficiency at higher electric powers.

For comparison, we carried out the same measurements of the AO diffraction efficiency for $TeO_2$ crystals as one of the best AO materials. The transverse acoustic wave with the frequency of 26.5 MHz was polarized along [001] and propagated in the direction [110], while the circularly polarized laser beam propagated along the direction [101]. It has been found that the value of AO diffraction efficiency for $TeO_2$ crystals is 2.3 times smaller than that for $Sn_2P_2S_6$ at the same power of the electric signal ($P_{el} = 1.56$ W). It can attain only the value of $\eta = 7\%$.

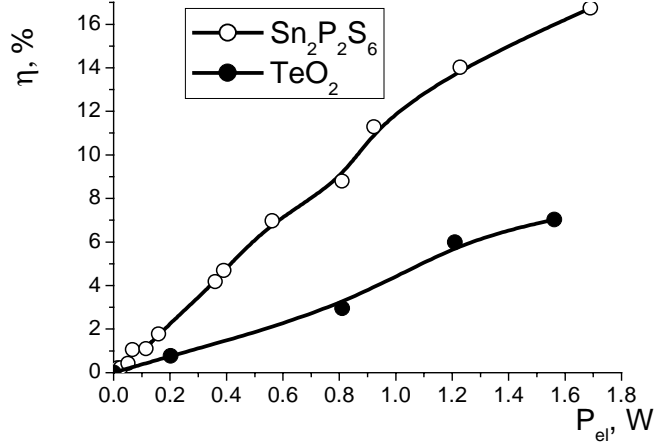

Figure 1. Dependences of AO diffraction efficiency on the electric driving power applied to the piezoelectric transducer that correspond to the laser beam (632.8 nm) interacting with the transverse acoustic wave in $Sn_2P_2S_6$ and $TeO_2$ crystals.

The diffraction efficiency may be described with the relation [9]

$$\eta = \sin^2\left(\frac{\pi}{\lambda \cos\theta}\sqrt{\frac{P_a \cdot L \cdot M_2}{2H}}\right), \qquad (2)$$

where $I_0$ denotes the intensity of the incident light, $\lambda$ the light wavelength, $\theta$ the diffraction angle, $P_a$ the acoustic power, $L$ the length of AO interaction, $M_2$ the AOFM and $H$ the width of the acoustic beam. For the case of weak interaction ($\eta \ll 1$) Eq. (2) may be simplified to the following form:

$$\eta = \frac{\pi^2 P_a \cdot L \cdot M_2}{2\lambda^2 \cos^2\theta \cdot H}. \qquad (3)$$

Basing on Eq. (3) and the value of AO diffraction efficiency obtained for $TeO_2$ crystals, we have calculated the acoustic power excited in the sample. It has turned out that the latter value is small enough, being equal to $P_a = 6.7\times10^{-4}$ W. Here the following parameter values have been used for the calculations: $\eta = 0.07$ (for the case of $P_{el} = 1.56$ W), $M_2 = 1200\times10^{-15}$ s$^3$/kg, $L = 7$ mm, $H = 1$ mm and $\lambda = 632.8$ nm. Since the experimental conditions of the measurements of AO diffraction efficiency for $Sn_2P_2S_6$ and $TeO_2$ crystals are the same, the acoustic power excited in both crystals should also be the same. Using Eq. (3) and the values $\eta = 0.16$ (at $P_{el} = 1.56$ W), $L = 5$ mm, $H = 1$ mm and $\lambda = 632.8$ nm, one can calculate the AOFM for $Sn_2P_2S_6$ crystals. It has been found that the AOFM for these crystals is even higher than we have expected (see [6,7]). Namely, the AOFM value is as high as $M_2 \simeq 3.8\times10^{-12}$ s$^3$/kg. The value of the effective photoelastic coefficient obtained on the basis of formula

$$|p_{ef}| = \sqrt{\frac{M_2 \rho v^3}{n^6}} \qquad (4)$$

and the data $\rho = 3.54 \times 10^3$ kg/m$^3$, $n = 3.0619$ [10], $v_{21} = 2100$ m/s and $v_{23} = 2610$ m/s [6] is equal to $|p_{ef}| \simeq 0.4 \div 0.5$. This is close to the values of photoelastic coefficients responsible for AO interactions, which have been calculated from the direct measurements of the piezooptic coefficients. It is worth noticing that, according to the data presented above, an Sn$_2$P$_2$S$_6$ crystal is the material that exhibits the highest value of the AOFM ever known for the visible spectral range.

**Conclusions**

In the present paper we have studied experimentally the AO diffraction occurring in Sn$_2$P$_2$S$_6$ crystals. It has been found that the AOFM for these crystals achieves a giant value, $M_2 \simeq 3.8 \times 10^{-12}$ s$^3$/kg. To our knowledge, this is the highest AOFM known for the AO materials operating in the visible spectral range. The consequence is that the acoustic powers as low as $P_a = 6.7 \times 10^{-4}$ W would be enough for gaining high diffraction efficiencies (e.g., $\eta = 16\%$).

**Acknowledgement**

The authors (M.O., M.-L.I. and V.R.) acknowledge financial support of this study from the Ministry of Education and Science of Ukraine (the Project N0106U000616).